# Main physical characteristics of crystal, magnetic and electronic structures of $Ce^{3+}$-based perovskites $CeTmO_3$ [$Tm^{3+}$ = Sc, Ti, V] investigated via the first-principles computational utilizing LDA, PBE-GGA and WC-GGA functionals

M. Musa Saad H.-E.

Department of Physics, College of Science and Arts in Muthnib, Qassim University, Muthnib 51931, Saudi Arabia; 141261@qu.edu.sa

## Abstract

Main physical characteristics such as crystal, magnetic and electronic structures of cerium-based perovskites $CeTmO_3$ compounds [$Tm^{3+}$ = Sc, Ti, V] are systematically investigated via the first-principles computations. Full-potential linear augmented plane-wave (FPLAPW) with the local density approximation (LDA) and generalized gradient approximation (GGA), under two functionals, Perdew-Burke-Ernzerh (PBE-GGA) and Wu-Cohen (WC-GGA), based on the density functional theory (DFT), executed in the Wien2k-19.1 package. The optimized crystal parameters indicate that the crystal of $CeTmO_3$ compounds can categorized under cubic structure (space group Pm-3m, no. 221), and their obtained lattice constants in good agreement with the available experimental values. The computed electronic band structures, density of states (DOS) and charge density reveal the metallic nature for the investigated $CeTmO_3$ compounds, except for [$Tm^{3+}$ = Sc] structure, exhibits half-metallic characteristics. Besides, LDA, PBE-GGA and WC-GGA functionals provide nearly similar results of crystal structures, DOSs as well as spin magnetic moments confirm that the $CeTmO_3$ compounds are ferromagnetic (FM) metals. The 2D and 3D electronic charge density designs in [100] plane indicate the existence of ionic bonding among ($O^{2-}$–$Ce^{3+}$–$O^{2-}$) and ($O^{2-}$–$Tm^{3+}$–$O^{2-}$) bonds to construct all $CeTmO_3$ crystals.

## 1. Introduction

Perovskites ($AMX_3$) and their related structures constitute an important family of magnetic materials and have received considerable attention for applications in solid-state physics, solid-state chemistry, advanced materials science and engineering materials due to their exceptional properties. Based on the types of constructed atoms, perovskites can be classified into three main groups; inorganic perovskites ($BaTiO_3$, $APbBr_3$) [1,2], organic perovskites ($PbNH_4Br_3$) [3] and inorganic-organic perovskites ($CH_3NH_3PbI_3$) [4]. The most studied group within this family in the current research studies is the inorganic metal perovskite oxides ($AMO_3$). Where, $AMO_3$ compounds (that containing two different metals; $A^+$-$M^{5+}$, $A^{2+}$-$M^{4+}$ or $A^{3+}$-$M^{3+}$) with varied chemical compositions and crystal structures have been investigated by many researchers. They are mainly motivated by their distinct structures and possible applications in materials science, manufacturing, engineering and other modern fields [5-10].

Among inorganic perovskites, rare-earth (RE) perovskites with universal chemical formula (REMO$_3$) form an important and promising subfamily of functional compounds. Therefore, in recent years, several members of metal perovskites based on these parent compound REMO$_3$ have been synthesized and studied by selecting its cation RE as follow: RE = La [5,6,11], Ce [10,12,13], Pr [14,15,16], Nd [11,17], Pm [18], Sm [19,20], Eu [21], Gd [22,23], Tb [22,24], Dy [22,25], Ho [26], Er [27], Tm [28], Yb [29] and Lu [30]. From these studies, a few authors are interested in exploring the chemical and physical properties of a major class of REMO$_3$, namely rare-earth aluminates REAlO$_3$ (RE$^{3+}$ = Pr, Nd, Lu) due to their utilizations in several applications, such as in dielectrics, ferroelectrics, magneto-optics, photocatalytic, laser devices and scintillator [10,16]. Most research groups focused their attention on studying the major characteristics of rare-earth transition-metal (Tm) perovskites. For instance, they found that some RETmO$_3$ compounds with (RE$^{3+}$= La, Pr, Nd; Tm$^{3+}$ = 3d-transition-metal) like LaTmO$_3$, PrTmO$_3$ and NdTmO$_3$ are considered significant materials for many novel technical applications, such as spintronics, fuel cells, chemical reactors and catalytics [5,6,11,12].

Many authors studied the magnetic, electronic, thermoelectric, ferroelectric and multiferroic properties in some rare-earth mangnaites REMnO$_3$. They found that the majority compounds of REMnO$_3$ crystallize in a cubic structure (Pm-3m; No. 221, space-group) when (RE = La, Ce, Pr) [12,15,31], and in an orthorhombic structure (Pnma or Pbnm; No. 62, space-group) when (RE = La, Nd, Sm, Gd, Tb, Dy, Tm Lu) [20,22,24,25,30,32,33,34], whereas a hexagonal structure has maintained in (P6$_3$cm; No. 185, space-group) within (RE = Ho, Yb, Ln) [25,35,36]. Accordingly, these properties suggest REMnO$_3$ compounds may be potential candidate for many novel applications, such as solid oxide fuel cell cathode, piezoelectric transducers, magnetic refrigerant, multilayered capacitors, computer memories and pyroelectric sensors [31,32,33].

Thus, from the above exposes, the physical properties of all RETmO$_3$ originate mainly from an interplay between their crystal structure type, chemical composition, as well as, internal factors like temperature, pressure and exchange-correlation (XC) energy. In this regard, the effect of some of these features on the physical properties of some 3d-transition-metal perovskite oxides RETmO$_3$ have been explored by some authors via utilizing certain computational techniques based on density functional theory (DFT) [12,15,31,38]. Where their DFT-based functionals under local density approximation (LDA) and/or generalized gradient approximation (GGA) achieved remarkable success in describing the crystal, electronic, magnetic and optical structures of REXO$_3$ compounds.

The current manuscript presents a first-principles investigation of the main physical characteristics such as crystal, magnetic and electronic structures of the cerium-based perovskites CeTmO$_3$ compounds [Tm$^{3+}$ = Sc, Ti, V] by utilizing the different possibilities of full-potential linear augmented plane-wave (FPLAPW) method. Three DFT functionals, LDA, Perdew-Burke-Ernzerh GGA (PBE-GGA) and Wu-Cohen GGA (WC-GGA), are optioned for the XC potential in all CeTmO$_3$ systems. To the best of our knowledge, this is the first systematical and comprehensive study on a new series inclosing rare-earth perovskites CeTmO$_3$ [Tm$^{3+}$ = Sc, Ti, V] that have an ideal crystal and electronic structures and excellent chemical stability simultaneously. Besides, CeTmO$_3$ series has not been studied before, either by using experimental techniques or theoretical methods.

The manuscript is organized as follow: in Section 2, the primary computational methodology, DFT calculations and their designing details are defined. In Section 3, the main results of physical characteristics that obtained for the concerned CeTmO$_3$ compounds are listed and discussed. Finally, the core conclusions of this manuscript are provided in Section 4.

## 2. Computational methodology and calculations details

In this study, first-principles computations of the main physical properties of CeTmO$_3$ [Tm$^{3+}$ = Sc, Ti, V] were performed by using the full-potential linear augmented plane wave (FPLAPW) method as executed in the Wien2k package [39]. The XC potential in these CeTmO$_3$ compounds was treated systematically by selecting the LDA, PBE-GGA and WC-GGA approximations within the DFT [39-41]. In order to obtain accurate results, the parameter ($R_{MT}K_{max}$ = 9.0) was used, which rules the convergence process, where $R_{MT}$ = the muffin-tin (MT) radii and $K_{max}$ = the plane-wave cutoff. Therefore, the values of $R_{MT}$ for all atoms in CeTmO$_3$ were chosen as ($R_{Ce}$ = 2.50, $R_{Tm}$ = 1.86 – 1.74 and $R_O$ = 1.69 – 1.57 a.u.), where these $R_{MT}$ insure that there is no any charges leakage from the internal atomic cores as well as to acquire exact energy eigenvalues convergency. Within these MTs, the potential V(r) and charge density ϱ (r) are expanded in terms of crystal harmonic up to angular momenta ($L_{max}$ = 10.0) and plane-wave expansion has been used in the interstitial region. Also, the Fourier expansion parameter set as ($G_{max}$ = 14.0), which delimits the magnitude of largest vector in ϱ (r). The Monkorst-Pack k-points were carried out by special (k-points = 1000) in the Brillouin zone (BZ). The cutoff energy was adapted as ($E_{cutoff}$ = 6.0 Ry) to describe well the separation energy between valence and core electrons in all atoms. For all CeTmO$_3$ compounds, the energy and charge convergence were fixed at the values ($E_{Conv.}$ = 0.0001 Ry) and ($ϱ_{Conv.}$ = 0.0001e) during the self-consistency cycles.

## 3. Results and discussion

In the current section, the main computations results of LDA, PBE-GGA and WC-GGA functionals for the crystal, magnetic and electronic structures of the three cerium-based perovskites CeTmO$_3$ compounds with [Tm$^{3+}$ = Sc, Ti, V] are presented separately in the following subsections and discussed in details.

### 3.1. Crystal structure and ground states

As an initial step, the ground states of perovskite unit cells CeTmO$_3$ [Tm$^{3+}$ = Sc, Ti, V] are examined via computing the total energy (E$_{Total}$) versus different values of the unit cell volume (V) around the equilibrium volume ($V_0$) by using LDA, PBE-GGA and WC-GGA functionals. Fig. 1 shows the three different plots of the optimized curves of the variation of E$_{Total}$ vs. unit cell V for CeTmO$_3$ compounds. Then, all the LDA, PBE-GGA and WC-GGA curves of E$_{Total}$ vs. V (represent by solid lines) are fitted to the Birch-Murnaghan equation of state, (denote by + symbols), as follows:

$$E(V) = E_0 + \frac{B_0 V}{B_0'}\left[\frac{(V_0/V)^{B_0'}}{(B_0'-1)} + 1\right] - \left[\frac{B_0 V_0}{(B_0'-1)}\right] \quad (1)$$

Table 1 summarizes the main obtained values of ground state energy ($E_0$), optimized parameters ($a_0$), equilibrium volume ($V_0$), bulk modulus ($B_0$) and first pressure derivative bulk modulus ($B_0'$) for the present compounds CeTmO$_3$ [Tm$^{3+}$ = Sc, Ti, V].

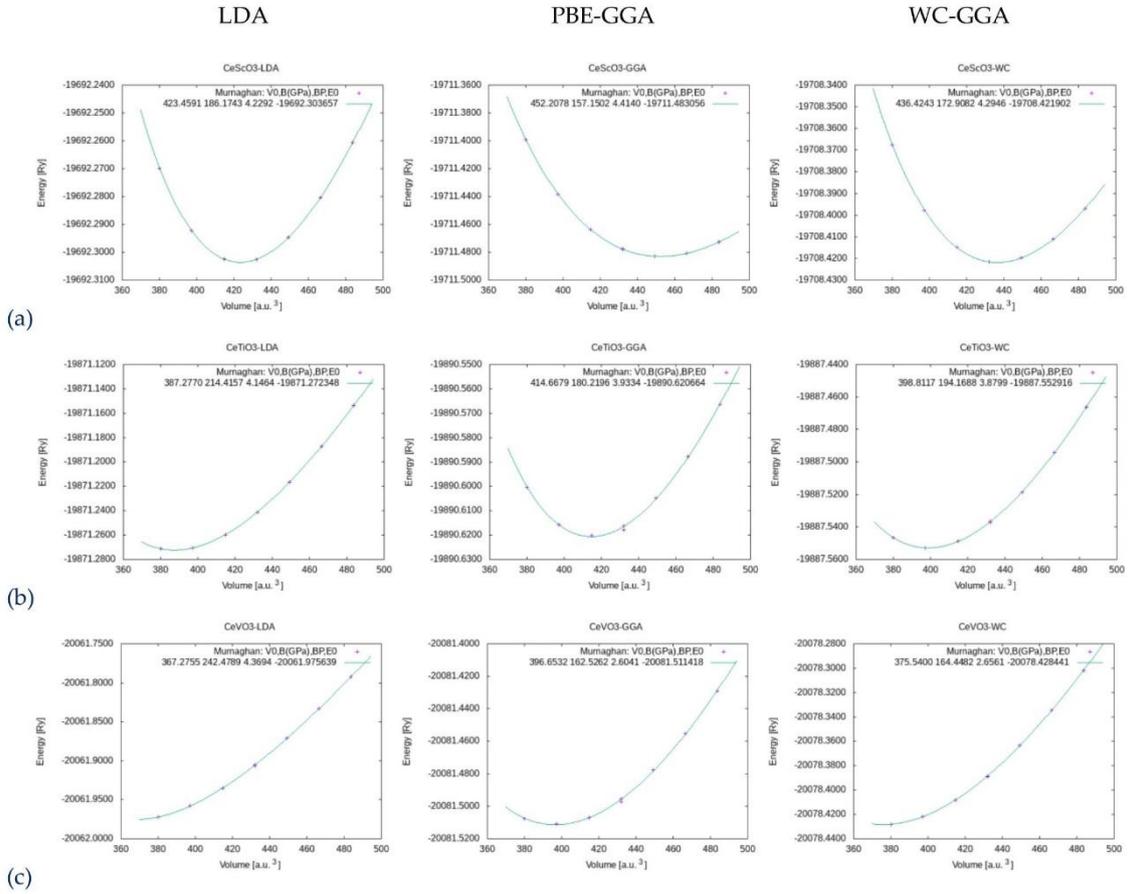

**Fig. 1.** Plots of the variation of total energy ($E_{Total}$) verses unit cell volume (V) for CeTmO$_3$ [Tm$^{3+}$ = Sc, Ti, V] within LDA, PBE-GGA and WC-GGA functionals.

**Table 1.** Main computed structural results, lattice constants ($a_0$), unit cell volumes ($V_0$), bulk modulus ($B_0$), derivative $B_0$ ($B_0'$) and ground state energy ($E_0$) for the cubic CeTmO$_3$ [Tm$^{3+}$ = Sc, Ti, V] perovskite compounds within (Pm-3m) symmetry using LDA, PBE-GGA and WC-GGA functionals.

| CeTmO$_3$ | FPLAPW functional | $a_0$ (Å) | $V_0$ (Å$^3$) | $B_0$ (GPa) | $B_0'$ (GPa) | $E_0$ (keV) |
|---|---|---|---|---|---|---|
| CeScO$_3$ | LDA | 3.9738 | 62.750 | 186.17 | 4.2292 | -267.93 |
| | PBE-GGA | 4.0618 | 67.010 | 157.15 | 4.4140 | -268.19 |
| | WC-GGA | 4.0139 | 64.6715 | 172.91 | 4.2946 | -268.15 |
| CeTiO$_3$ | LDA | 3.8572 | 57.389 | 214.42 | 4.1464 | -270.36 |
| | PBE-GGA | 3.9461 | 61.448 | 180.22 | 3.9334 | -270.63 |
| | WC-GGA | 3.8952 | 59.098 | 194.169 | 3.8799 | -270.58 |
| CeVO$_3$ | LDA | 3.7896 | 54.425 | 242.48 | 4.3694 | -272.96 |
| | PBE-GGA | 3.8881 | 58.778 | 162.53 | 2.6041 | -273.22 |
| | WC-GGA | 3.8179 | 55.649 | 164.448 | 2.6561 | -273.18 |

From the above results, it can be concluded that all three crystal structures of CeTmO$_3$ compounds exhibit cubic symmetry (space group Pm-3m; # 221) with lattice constants $a = b = c = a_0$ ($a_0$ = ~3.80 – 4.06 Å) and $\alpha = \beta = \gamma = 90°$, in close agreement with the experimental and previous DFT results [42,43]. Also, it can be clearly observed that there are slight differences between all the optimized parameters obtained by LDA, PBE-GGA and WC-GGA functionals, and that LDA gives the smaller results. Commonly, the values of lattice constants of perovskite crystals CeTmO$_3$ as well as their unit cell volumes decrease linearly in going from Tm = Sc to Tm = V compounds with all DFT functionals. While, the bulk parameters $B_0$ and B' increase in this sequence, which indicates that the compound containing Tm = V has a harder nature and less compressible than that for other two compounds with Tm = Ti and Tm = Sc. Furthermore, the increasing in ground state energy ($E_0$), which gives an evidence for the crystal structure stability of these compounds, depends mainly on the replacing of Tm site; $E_0 (CeScO_3) > E_0 (CeTiO_3) > E_0 (CeVO_3)$. Thus, the ionic radius of Tm site (Table 2) governors these variations in $E_0$ value.

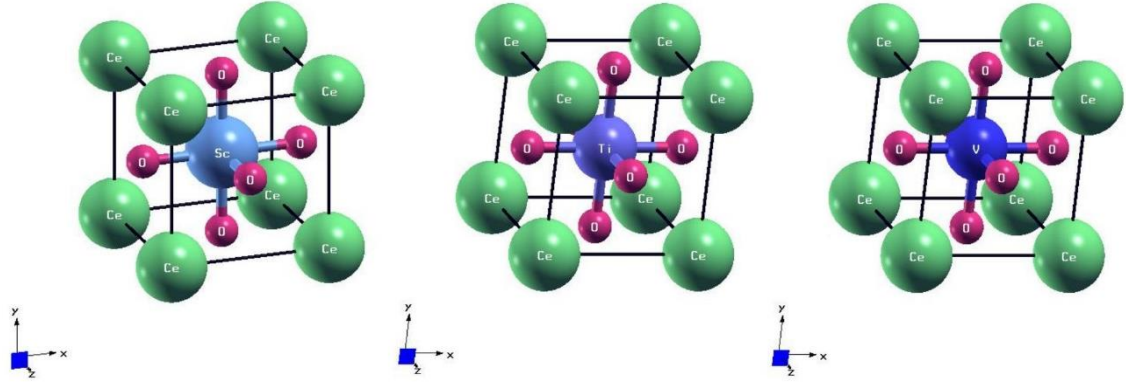

**Fig. 2.** Crystal structures of the cubic (Pm-3m) unit cells for perovskites CeTmO$_3$, where the green spheres at the lattice corners are Ce$^{3+}$-cations, the blue sphere at the center is Tm$^{3+}$-cation [Tm$^{3+}$ = Sc, Ti, V] and the pink spheres at the lattice faces are O$^{2-}$-anions.

Fig. 2 displays the optimized crystal structures of the cubic (Pm-3m) unit cells for CeTmO$_3$ [Tm$^{3+}$ = Sc, Ti, V]. The crystal creation of the compounds under this study CeTmO$_3$ can be conceived as six O$^{2-}$ ions surround the Tm$^{3+}$ ions which occupy the central position of the unit cell (0.5, 0.5, 0.5). These nearby O$^{2-}$ ions are situated at the faces of the unit cell at (0.5, 0.5, 0) position, whereas the Ce$^{3+}$ ions take up the corners at (0, 0, 0). Table 2 shows the ionic radii, Ce$^{3+}$ in 12- and Tm$^{3+}$ and O$^{2-}$ in 6- coordination systems, and conventional atomic Wyckoff positions and coordinates of Ce, Tm and O sites in the general cubic unit cells of CeTmO$_3$.

**Table 2.** Ionic radius and general atomic positions in cubic crystal structure (Fig. 2) of the unit cells for CeTmO$_3$ [Tm$^{3+}$ = Sc, Ti, V]

| Site | R (Å) | Wyckoff position | Atomic coordinates | | |
|---|---|---|---|---|---|
| | | | x | y | z |
| Ce-cation | 1.34 | 1a | 0 | 0 | 0 |
| Tm-cation | Sc = 0.745, Ti = 0.67, V = 0.64 | 1b | ½ | ½ | ½ |
| O-anion | 1.40 | 3c | ½ | ½ | 0 |
| | | | ½ | 0 | ½ |
| | | | 0 | ½ | ½ |

One more crystal structure parameters are the bond-lengths and bond-angles, which have their importance to determine the symmetry of the perovskite unit cells CeTmO$_3$ [Tm$^{3+}$ = Sc, Ti, V]. Where the atomic electronic configurations in these perovskites take the ordinary forms; Ce [Xe] 6s$^2$ 4f$^1$ 5d$^1$, Sc [Ar] 4s$^2$ 3d$^1$, Ti [Ar] 4s$^2$ 3d$^2$, V [Ar] 4s$^2$ 3d$^3$ and O [He] 2s$^2$ 2p$^4$. The values of muffin-tin radii (R$_{MT}$) and main bond-lengths Ce$^{3+}$–O$^{2-}$, Tm$^{3+}$–O$^{2-}$ and Ce$^{3+}$–Tm$^{3+}$ are computed and listed in Table 3 within these DFT functionals. Here, Ce$^{3+}$–O$^{2-}$ = Tm$^{3+}$–O$^{2-}$ = Ce$^{3+}$–Tm$^{3+}$ and depend only on the value of $a_0$, also, the main bond-angles are similar, <Ce$^{3+}$–O$^{2-}$> = 90º, <Tm$^{3+}$–O$^{2-}$> = 90º and <O$^{2-}$–Tm$^{3+}$–O$^{2-}$> = 180º, in all three cubic compounds, regardless of the functional operated in DFT calculations. Furthermore, these bond-lengths are used to estimate another crystal structure parameter, namely, tolerance factor (T$_F$) through the following ratio equation:

$$T_F = \frac{\langle R_{Ce}+R_O \rangle}{\sqrt{2}\langle R_{Tm}+R_O \rangle} \qquad (2)$$

The perovskite structure requires that the cations $Ce^{3+}$ and $Tm^{3+}$ occupy distinct coordination sites in the crystal unit cells of $CeTmO_3$; $Ce^{3+}$ (12-fold) and $Tm^{3+}$ (6-fold) coordinated by anions $O^{2-}$ to form stable tetrahedra $[Ce^{3+}O_{12}]$ surrounded by corner-sharing octahedra $[Tm^{3+}O_6]$ network. In the above equation, the $T_F$ is used to predict the crystal structure stability of perovskites $CeTmO_3$ based only on their chemical composition and ionic radii (R) of $Ce^{3+}$, $Tm^{3+}$ = Sc, Ti, V and $O^{2-}$ ions, as shown in Table 2. The computed values of $T_F$ for $CeTmO_3$ compounds are displayed in Table 3. From these results, it is concluded that the achieved values of $T_F$ have a close agreement with those calculated in previous studies for other rare-earth cubic perovskites, $RrXO_3$ with X = V, Cr, Mn and Fe [15,44]. The expectable value of $T_F$ for cubic perovskites is among the range 0.90 – 1.03 [44,45]. Accordingly, the current values of $T_F$ lie in this range, which confirm the cubic crystal structure (Pm-3m) of concerned $CeTmO_3$ compounds.

**Table 3.** The computed muffin-tin radii, tolerance factor and the main bond-lengths in the cubic unit cells of $CeTmO_3$ [$Tm^{3+}$ = Sc, Ti, V]

| CeTmO₃ | FPLAPW functional | $R_{MT}$ | | | $T_F$ | Tm–O (Å) |
|---|---|---|---|---|---|---|
| | | $Ce^{3+}$ | $Tm^{3+}$ | $O^{2-}$ | | |
| CeScO₃ | LDA | 2.50 | 1.82 | 1.65 | 0.9054 | 1.9869 |
| | PBE-GGA | 2.50 | 1.86 | 1.69 | | 2.0309 |
| | WC-GGA | 2.50 | 1.84 | 1.67 | | 2.0070 |
| CeTiO₃ | LDA | 2.50 | 1.77 | 1.60 | 0.9360 | 1.9286 |
| | PBE-GGA | 2.50 | 1.81 | 1.64 | | 1.9730 |
| | WC-GGA | 2.50 | 1.79 | 1.62 | | 1.9476 |
| CeVO₃ | LDA | 2.50 | 1.74 | 1.57 | 0.9497 | 1.8948 |
| | PBE-GGA | 2.50 | 1.78 | 1.61 | | 1.9441 |
| | WC-GGA | 2.50 | 1.75 | 1.58 | | 1.9089 |

### 3.2. Magnetic structure of ground states

Due to their exclusive electronic structures, it is very important to investigate the magnetic structures of rare-earth based perovskites $CeTmO_3$ because apart from $Tm^{3+}$ ions, it is expected that the $Ce^{3+}$ ions will contribute a large amount to the total spin magnetic moments of the unit cell. The ionic electronic configurations in $CeTmO_3$ take the energetic forms; $Ce^{3+}$ [Xe] $4f^1$, $Sc^{3+}$ [Ar] $3d^0$, $Ti^{3+}$ [Ar] $3d^1$, $V^{3+}$ [Ar] $3d^2$ and $O^{2-}$ [He] $2s^2\ 2p^6$. Table 4 summarizes the computed partial ($M_{Ce^{3+}}$, $M_{Tm^{3+}}$ and $M_{O^{2-}}$), interstitial ($M_{Int}$) and total spin magnetic moments ($M_{Cell}$) of the cubic unit cell for $CeTmO_3$ [$Tm^{3+}$ = Sc, Ti, V] by using the three DFT functionals, LDA, PBE-GGA and WC-GGA. As an initial probe, it can be observed that there are minor differences between the results of these functionals, and that LDA yields the slighter spin magnetic moments. Also, it can be seen from these values, the partial spin magnetic moments of $Ce^{3+}$ and $Tm^{3+}$ ions contribute by the majority part of the total spin magnetic moments ($M_{Cell}$) in all CeTmO$_3$ unit cell. Within all functionals, the total spin magnetic moment of CeTmO$_3$ unit cells increases from Tm$^{3+}$ = Sc to Ti and V, indicating the effect of number of unpaired electrons in 3d orbitals. This can be confirmed by Hund's rules, which require greater total spin magnetic moment in CeTmO$_3$ with Tm$^{3+}$ = V as compared to Tm$^{3+}$ = Ti and Sc. Moreover, the strong hybridizations between the different ions and interstitial sites in CeTmO$_3$ reduce the partial spin magnetic moments carried by the magnetic ions of Ce$^{3+}$ and Tm$^{3+}$ from their ordinary values and transferred fraction of spin magnetic moments to nonmagnetic O$^{2-}$ anions ($M_{O^{2-}}$) and interstitial sites ($M_{Int}$). The positive sign on spin magnetic moments of cations, Ce$^{3+}$ and Tm$^{3+}$, and negative sign on anions, O$^{2-}$, indicate the antiparallel exchange interaction of the spins alignment. In CeTmO$_3$ compounds, the total spin magnetic moment is enhanced within all DFT functionals, and $M_{Ce^{3+}}$ and $M_{Tm^{3+}}$ favor the parallel alignment of the spin magnetic moments, which is dominant due to the ferromagnetic (FM) exchange interaction between these spins. Whereas, $M_{O^{2-}}$ favor the antiparallel alignment of the spin magnetic moments. Strikingly, the total spin magnetic moment of [Tm$^{3+}$ = Sc] compound is ($M_{Cell} \approx$ 1.00 $\mu_B$), and integer value evidences the half-metallic ferromagnetic (HM-FM) feature in first compound.

**Table 4.** The computed partial, interstitial and total spin magnetic moments ($\mu_B$) in the cubic unit cell of CeTmO$_3$ [Tm$^{3+}$ = Sc, Ti, V] using LDA, PBE-GGA and WC-GGA functionals.

| CeTmO$_3$ | FPLAPW functional | $M_{Ce^{3+}}$ | $M_{Tm^{3+}}$ | $M_{O^{2-}}$ | $M_{Int}$ | $M_{Cell}$ |
|---|---|---|---|---|---|---|
| CeScO$_3$ | LDA | 0.9759 | 0.0040 | -0.0126 | 0.0589 | 1.0011 |
| | PBE-GGA | 0.9799 | 0.0028 | -0.0081 | 0.0422 | 1.0005 |
| | WC-GGA | 0.9740 | 0.0036 | -0.0070 | 0.0434 | 0.9999 |
| CeTiO$_3$ | LDA | 1.0741 | 0.0639 | -0.0059 | 0.1547 | 1.2750 |
| | PBE-GGA | 1.1762 | 0.0706 | -0.0057 | 0.1451 | 1.3749 |
| | WC-GGA | 1.1277 | 0.0811 | -0.0055 | 0.1636 | 1.3560 |
| CeVO$_3$ | LDA | 0.9270 | 0.1665 | -0.0043 | 0.1474 | 1.2280 |
| | PBE-GGA | 1.1018 | 1.1824 | -0.0016 | 0.4635 | 2.7428 |
| | WC-GGA | 0.9934 | 0.5424 | -0.0010 | 0.2718 | 1.8104 |

### 3.3. Electronic structure of ground states

The computed results of spin-up and spin-down electronic structures, band structure, density of states (DOS) and charge density, for the three concerned perovskites CeTmO$_3$ with [Tm$^{3+}$ = Sc, Ti, V] are displayed in the following Figs. 3 – 12. Figs. (3, 4, 5) show the band structures, whereas in Figs. (6, 7) and (8, 9, 10), the total density of states (TDOS) of CeTmO$_3$ and their partial density of states (PDOS) of Ce, Tm and O atoms are presented collectively for these compounds obtained from LDA, PBE-GGA and WC-GGA functionals, respectively. The 2D and 3D charge densities are plotted in Figs. (11, 12).

#### 3.3.1. Band structures

The computed spin-up and spin-down electronic band structures along the principle symmetry points of CeTmO$_3$ compounds are shown in Fig. 3 [Tm$^{3+}$ = Sc], Fig. 4 [Tm$^{3+}$ = Ti], and Fig. 5 [Tm$^{3+}$ = V] from LDA, PBE-GGA and WC-GGA functionals. In general, these plots clearly illustrate that the spin-up channels of all studied perovskites CeTmO$_3$ show a metallic nature while their spin-down channels are insulating for [Tm$^{3+}$ = Sc] compound, and conducting for [Tm$^{3+}$ = Ti] and [Tm$^{3+}$ = V] ones. As a result, the summed effect of both spin-up and spin-down channels in CeTmO$_3$ compounds within all three DFT functionals leads to a half-metallic (HM) property for the first compound [Tm$^{3+}$ = Sc] and a metallic property for the other two compounds [Tm$^{3+}$ = Ti, V]. Table 5 displays the computed energies of Fermi level (E$_F$) and values of spin-up and spin-down band gap (E$_g$), besides the resulting magnetic and electronic states in the cubic unit cell of CeTmO$_3$ due substitution of Tm$^{3+}$-site [Tm$^{3+}$ = Sc, Ti, V], parallels with the effect of application of LDA, PBE-GGA and WC-GGA functionals.

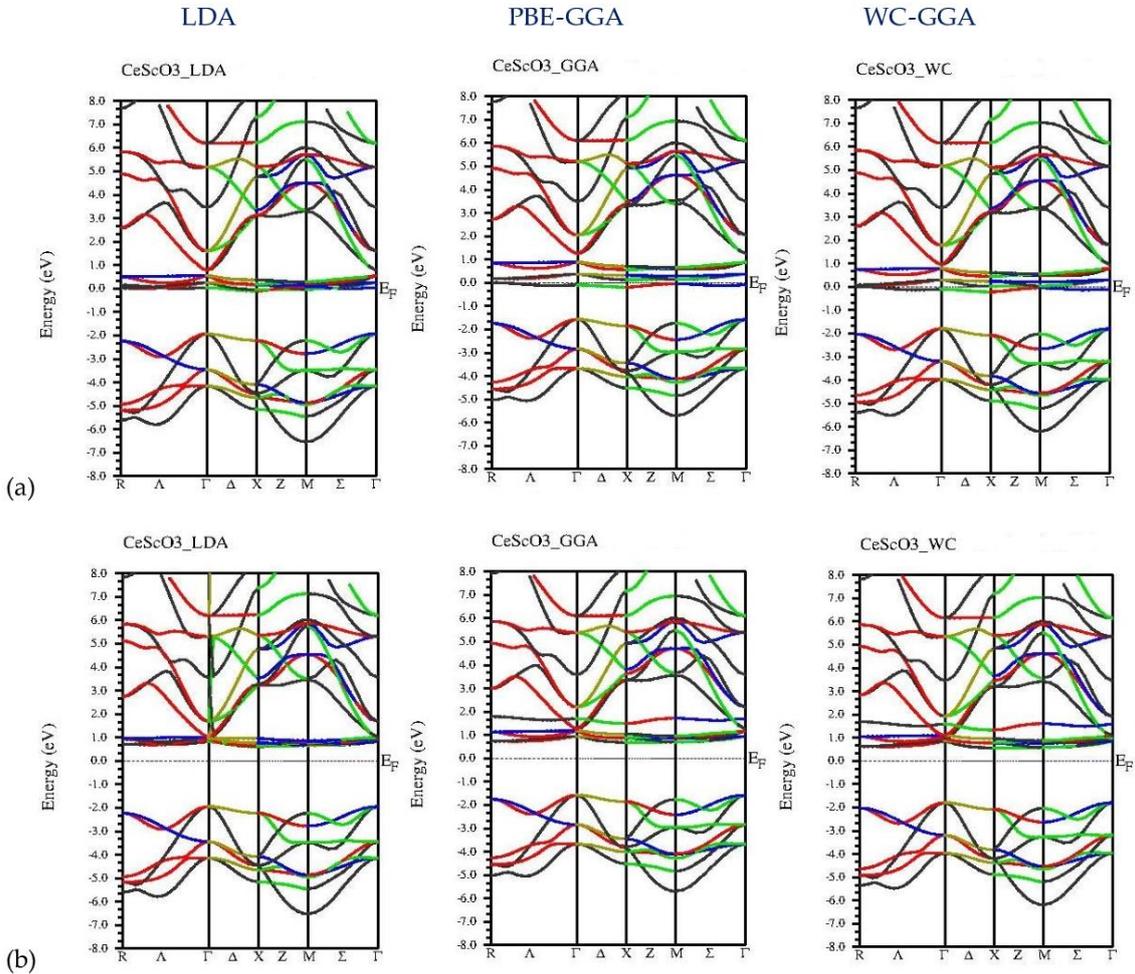

Fig. 3. Spin-up (a-panels) and spin-dn (b-panels) band structures for CeScO$_3$ obtained from LDA, PBE-GGA and WC-GGA functionals.

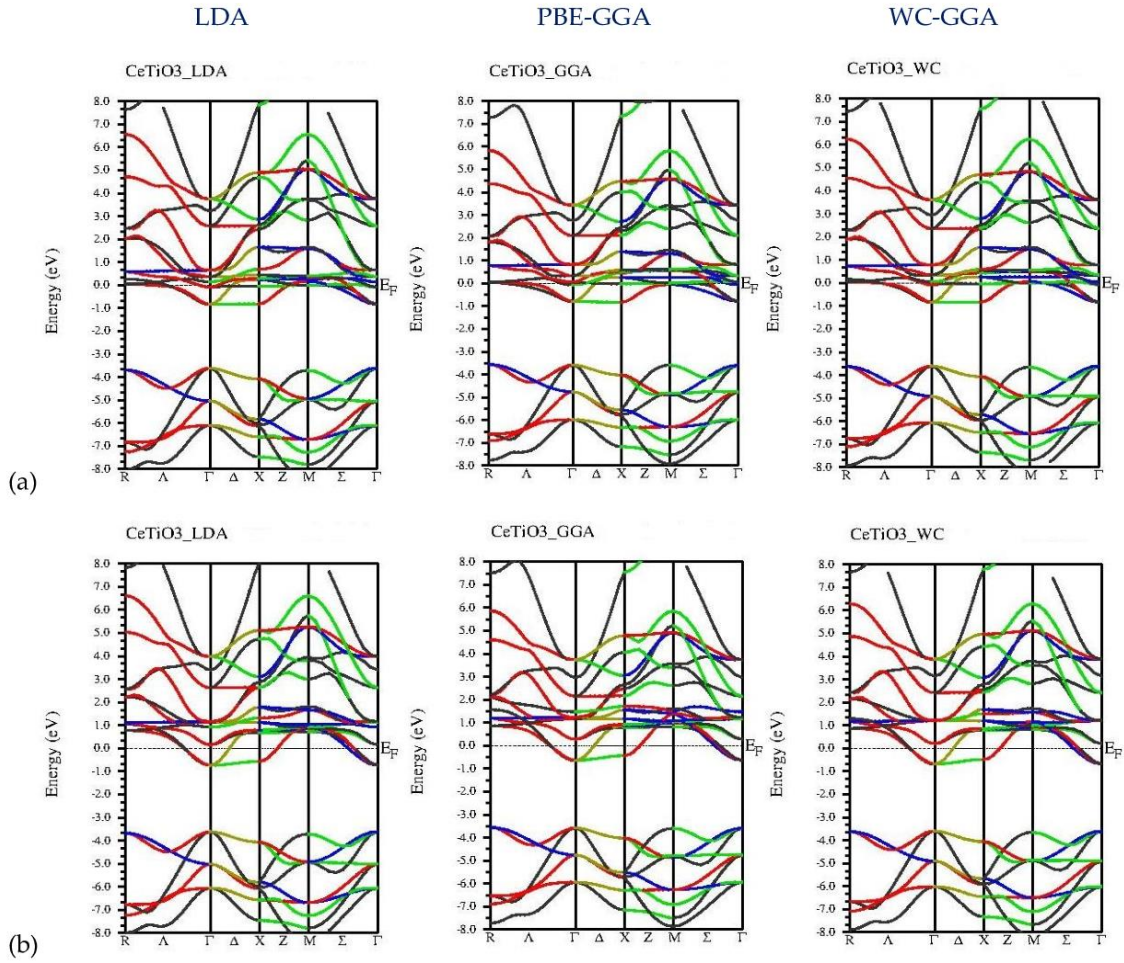

Fig. 4. Spin-up (a-panels) and spin-dn (b-panels) band structures for CeTiO$_3$ obtained from LDA, PBE-GGA and WC-GGA functionals.

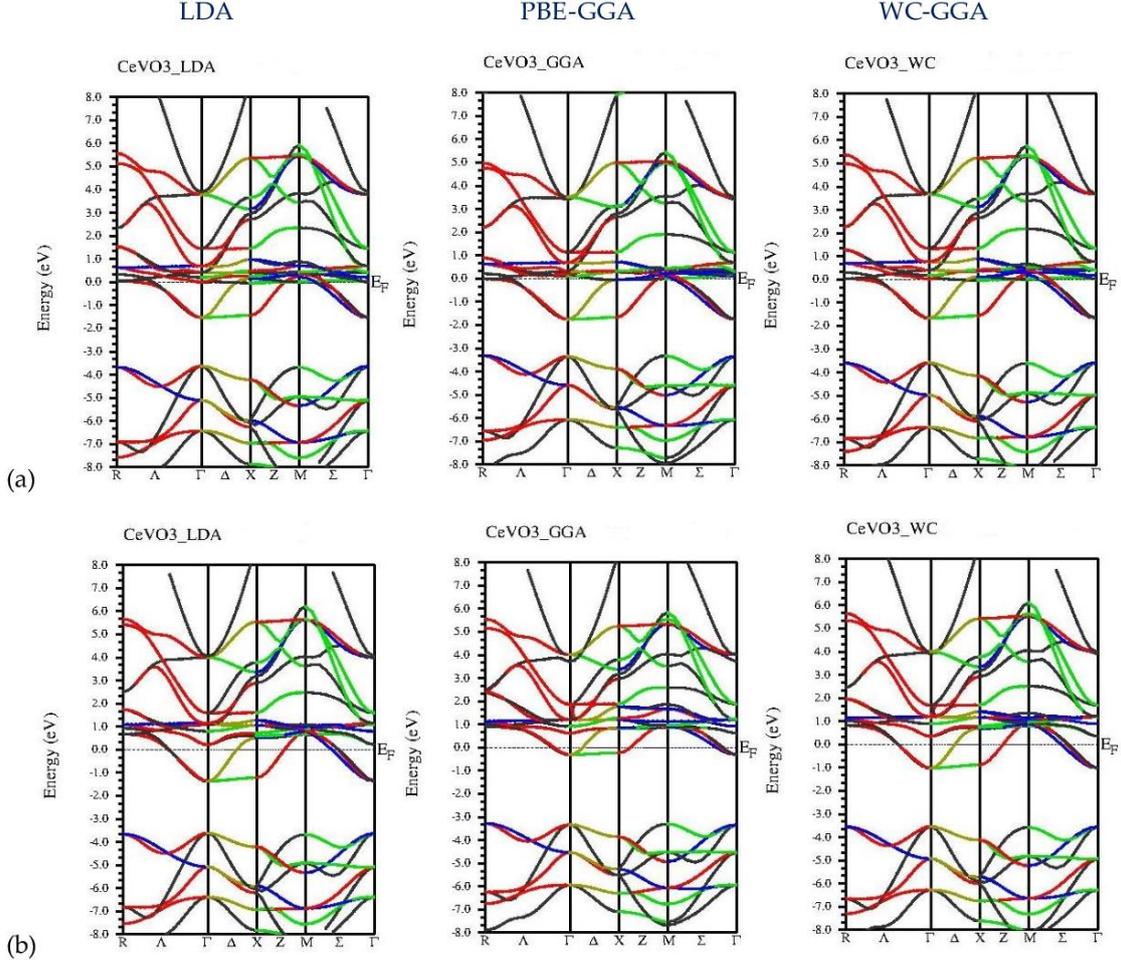

Fig. 5. Spin-up (a-panels) and spin-dn (b-panels) band structures for CeVO$_3$ obtained from LDA, PBE-GGA and WC-GGA functionals.

**Table 5.** The computed energies of Fermi level and values of spin-up and spin-down band gap, plus the obtained magnetic and electronic states in the cubic unit cell of CeTmO$_3$ [Tm$^{3+}$ = Sc, Ti, V] from LDA, PBE-GGA and WC-GGA functionals.

| CeTmO$_3$ | FPLAPW functional | $E_F$ (eV) | $E_{g\uparrow}$ (eV) | $E_{g\downarrow}$ (eV) | Magnetic State | Electronic State |
|---|---|---|---|---|---|---|
| CeScO$_3$ | LDA | 9.7502 | 0.0000 | 2.6440 | FM | HM |
|  | PBE-GGA | 8.8403 | 0.0000 | 2.3170 | FM | HM |
|  | WC-GGA | 9.3757 | 0.0000 | 2.4250 | FM | HM |
| CeTiO$_3$ | LDA | 12.096 | 0.0000 | 0.0000 | FM | M |
|  | PBE-GGA | 11.432 | 0.0000 | 0.0000 | FM | M |
|  | WC-GGA | 11.823 | 0.0000 | 0.0000 | FM | M |
| CeVO$_3$ | LDA | 12.610 | 0.0000 | 0.0000 | FM | M |
|  | PBE-GGA | 11.588 | 0.0000 | 0.0000 | FM | M |
|  | WC-GGA | 12.407 | 0.0000 | 0.0000 | FM | M |

### 3.3.2. *Total and partial density of states*

Similarly, the obtained electronic properties by band structures can be distinguished through the distributions of TDOS and PDOS for CeTmO$_3$ compounds as illustrated in Figs. 6 – 10. As it known that the conduction electrons in HM crystals, as in CeScO$_3$ in Fig. 6, should be 100% percent spin-polarized (SP), in spin-up (SP = +1) or spin-down (SP = -1) channel, at the Fermi level (E$_F$). The SP ration of CeTmO$_3$ can be estimated by means of spin-up DOS$^\uparrow$ and spin-down DOS$^\downarrow$ at E$_F$ through the relation:

$$SP = \frac{DOS^\uparrow - DOS^\downarrow}{DOS^\uparrow + DOS^\downarrow} \times 100\% \qquad (3)$$

Thus, the computed value of (SP = +1) at E$_F$ for CeScO$_3$, which obtained from (DOS$^\uparrow$ = +1) and (DOS$^\downarrow$ = 0), confirms the HM feature in CeScO$_3$ compound with a complete SP in spin-up direction. From Figs. 6 and 7, the HM band-gap (E$_g$) in this compound increases from LDA (E$_g$ = 2.644 eV) to PBE-GGA (E$_g$ = 2.317 eV) and WC-GGA (E$_g$ = 2.425 eV). The increasing value of the E$_g$ may be associated with the increasing number of electrons in the shell that increases the Coulomb repulsion effect, which shifts the partial states away from the E$_F$. Conversely, the spin-up and spin-down TDOS within the DFT functionals confirm the metallic property that obtained through the band structure computations for [Tm$^{3+}$ = Ti] and [Tm$^{3+}$ = V] compounds.

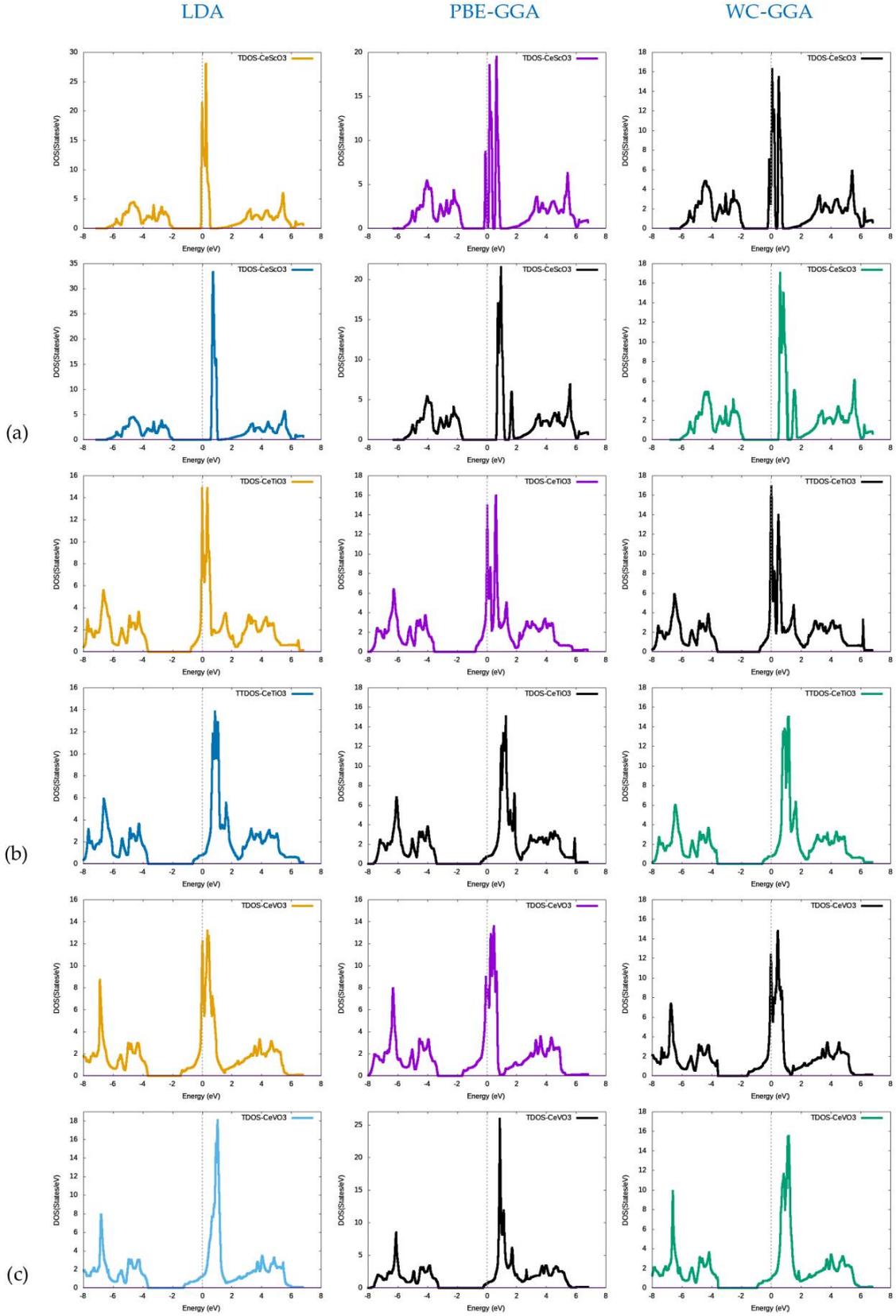

**Fig. 6.** Spin-up (up panels) and spin-dn (down panels) total density of states for CeTmO$_3$ [Tm$^{3+}$ = (a) Sc, (b) Ti, (c) V] obtained from LDA, PBE-GGA and WC-GGA functionals.

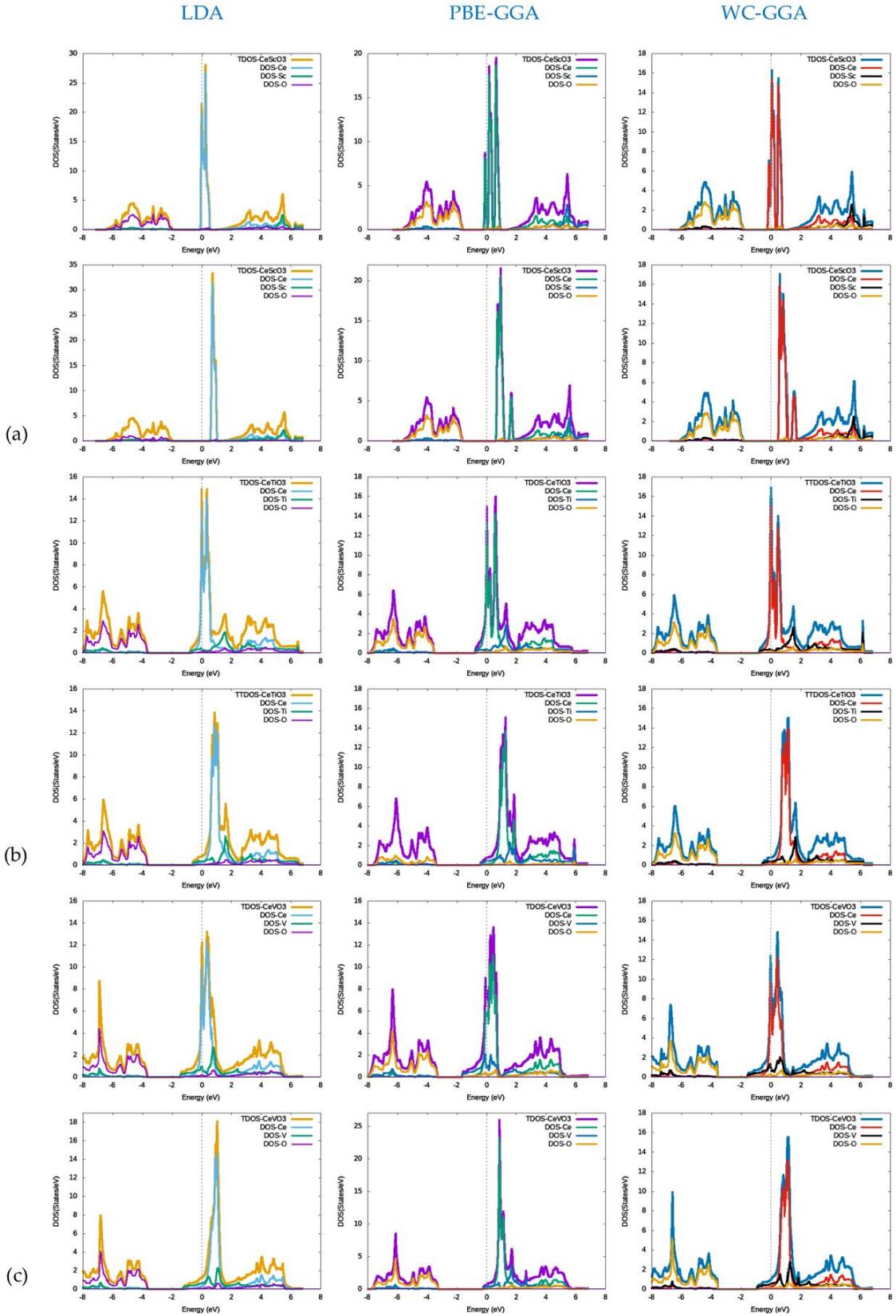

**Fig. 7.** Spin-up (up panels) and spin-dn (down panels) total and partial density of states, Ce, Tm and O, for CeTmO$_3$ [Tm$^{3+}$ = (a) Sc, (b) Ti, (c) V] obtained from LDA, PBE-GGA and WC-GGA functionals.

Moreover, to more elucidate the different contributions of the three ions $Ce^{3+}$, $Tm^{3+}$ and $O^{2-}$ in their $CeTmO_3$ compounds, the spin-up and spin-down PDOSs are computed via the LDA, PBE-GGA and WC-GGA functionals, and plotted separately in Fig. 8 [$Tm^{3+}$ = Sc], Fig. 9 [$Tm^{3+}$ = Ti] and Fig. 10 [$Tm^{3+}$ = V]. It is clear from the spin-up PDOS curves (up panels) there are orbital hybridizations around $E_F$ between the states of $CeTmO_3$; slight $Tm^{3+}$-3d–$O^{2-}$-2p in the energy range -1.60 eV to $E_F$, and considerable $Tm^{3+}$-3d–$Ce^{3+}$-4f in the range -0.45 eV to 0.85 eV. While in spin-down PDOS curves (down panels), the hybridizations $Tm^{3+}$-3d–$O^{2-}$-2p (-1.60 eV – 0.0 eV) and $Tm^{3+}$-3d–$Ce^{3+}$-4f (-1.60 eV – 0.0 eV) appear only in [$Tm^{3+}$ = Ti] and [$Tm^{3+}$ = V] compounds. Thus, these hybridizations confirm the partial contribution to electronic and magnetic structures of the $CeTmO_3$ compounds that come from $Ce^{3+}$ and $Tm^{3+}$ ions.

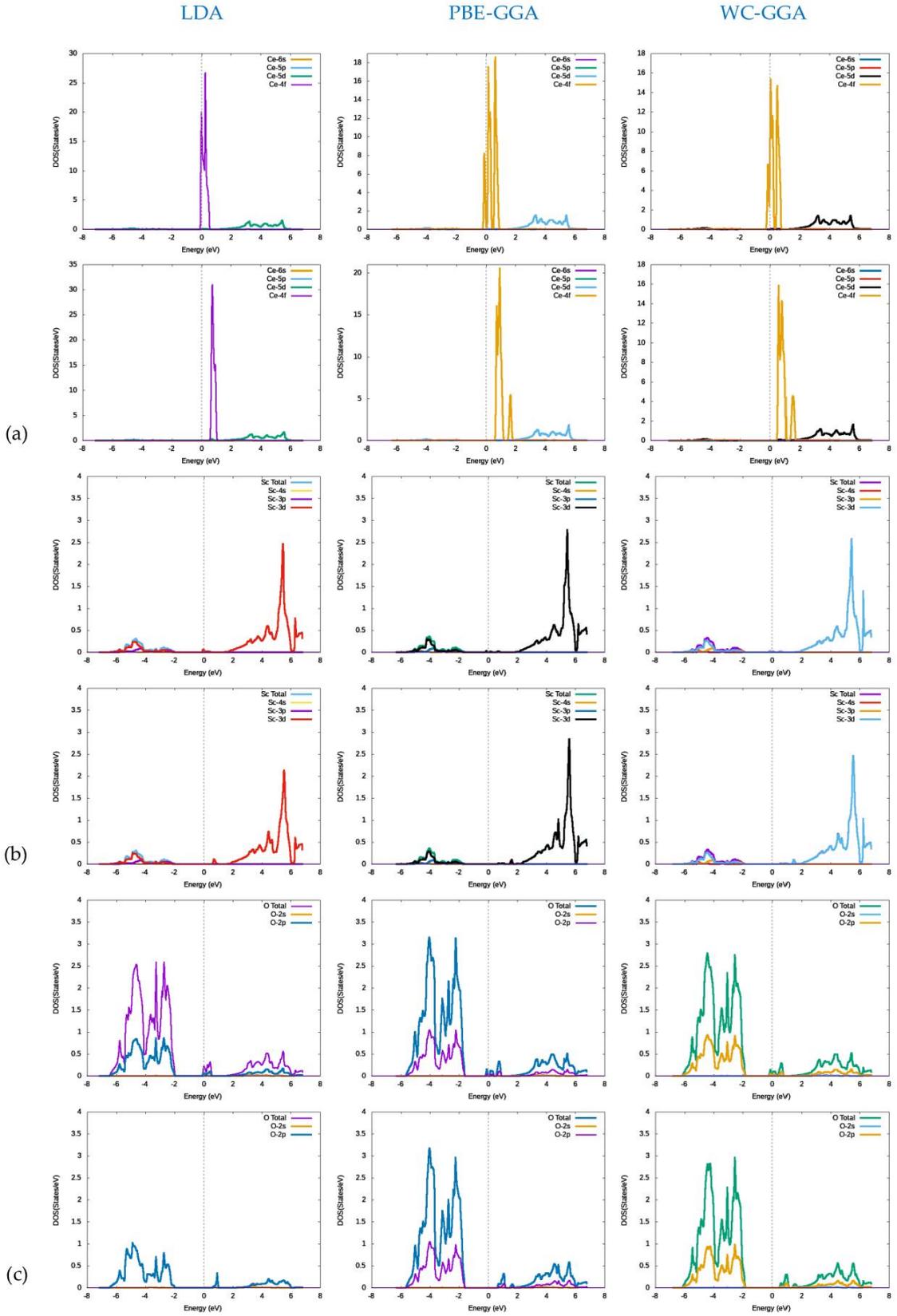

**Fig. 8.** Spin-up (up panels) and spin-dn (down panels) partial density of states of Ce, Sc and O atoms in CeScO$_3$ obtained from LDA, PBE-GGA and WC-GGA functionals.

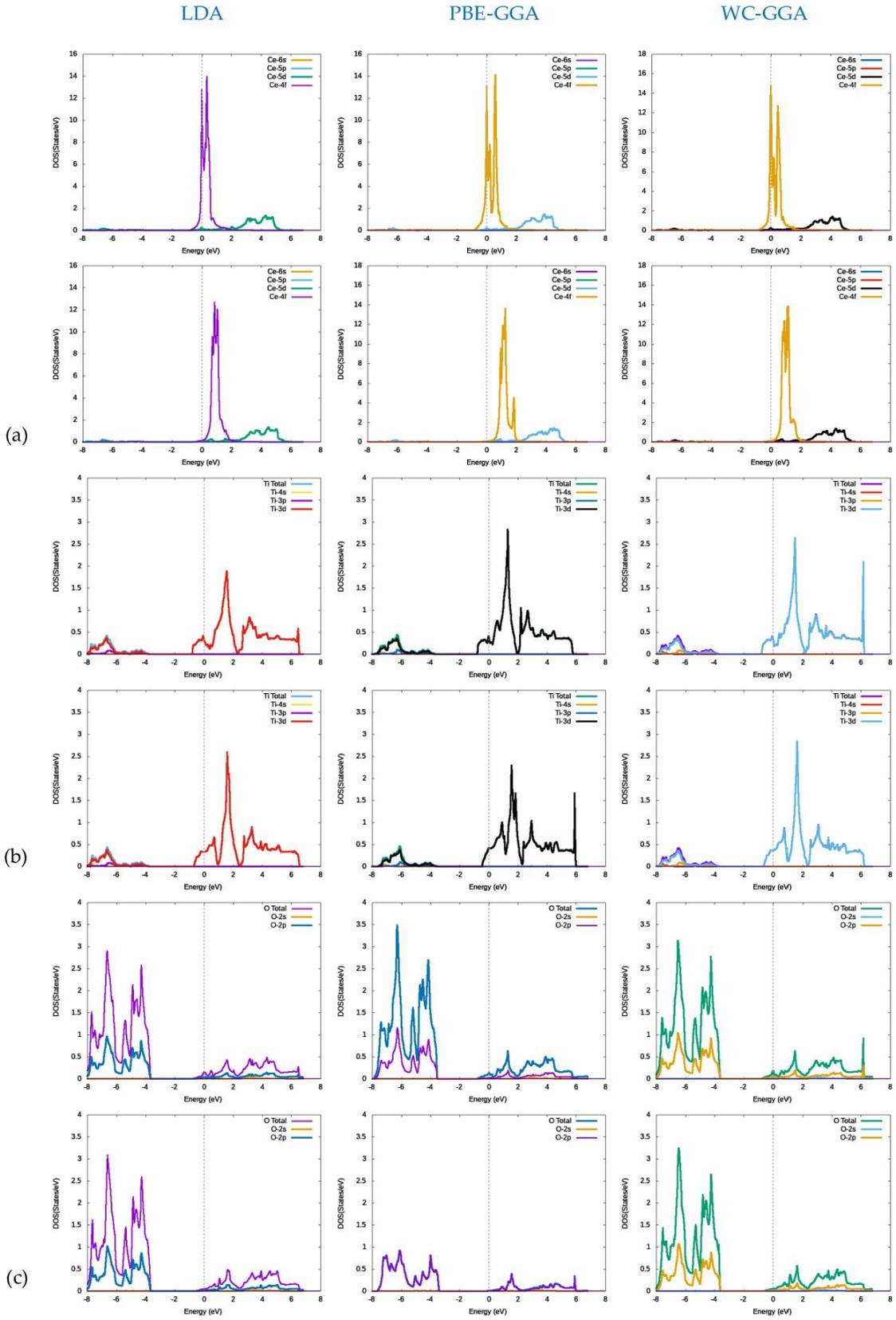

**Fig. 9.** Spin-up (up panels) and spin-dn (down panels) partial density of states of Ce, Ti and O atoms in CeTiO$_3$ obtained from LDA, PBE-GGA and WC-GGA functionals.

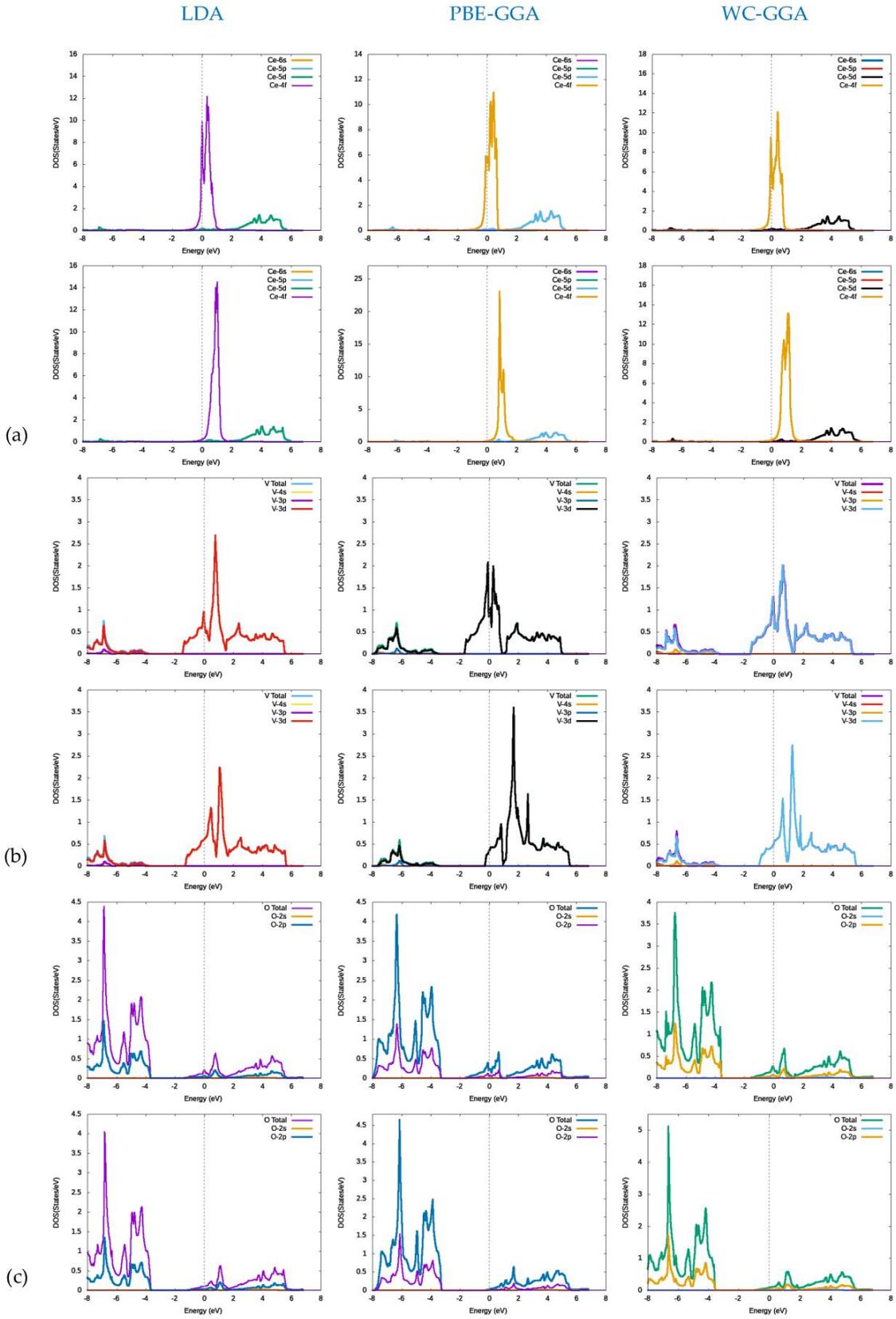

**Fig. 10.** Spin-up (up panels) and spin-dn (down panels) partial density of states of Ce, V and O atoms in CeVO$_3$ obtained from LDA, PBE-GGA and WC-GGA functionals.

### 3.3.3. *Electronic charge density and chemical bonding*

Generally, two dimensional (2D) and three (3D) plots of electronic charge density describe the chemical bonding nature in any structure of materials. Therefore, useful information can be obtained from these plots about the transfer of charges ($\Delta\rho_z$) or/and charges sharing between cations and anions, which are used to distinguish the pattern and nature of ionic and covalent bonding in crystalline materials. Fig. 11 shows the 2D contour plots of the total electronic charge density along [100] plane for the three perovskites $CeTmO_3$ [$Tm^{3+}$ = Sc, Ti, V], obtained from LDA, PBE-GGA and WC-GGA functionals. Besides interstitial sites, all plots include the energetic states of cations, Ce-4f and Tm-3d, and anions O-2p, whose orbitals can be distinguished through the shapes of contour lines. From these designs, the chemical bonding contains partially ionic and partially covalent in nature with electronic charge density around the anions $O^{2-}$ for both spin-up and spin-down directions. It can be seen the cations $Ce^{3+}$ and anions $O^{2-}$ form the covalent bonds ($O^{2-}$–$Ce^{3+}$–$O^{2-}$) in all these crystals, where the maximum electronic charge density, represented by condensed contour lines distributions, exists around to the sites of both $Ce^{3+}$ and $O^{2-}$ ions. Whereas, the pattern of bonds between the other cations $Tm^{3+}$ = Sc, Ti, V, at the center, and anions $O^{2-}$, in [$Tm^{3+}O^{2-}_6$] octahedra, are purely ionic bonds ($O^{2-}$–$Tm^{3+}$–$O^{2-}$) with electronic charge density that sharing among the contour lines of these sites. Moreover, the contour lines distributions take the spherical shape in the cations $Ce^{3+}$ and anions $O^{2-}$, which indicates the ionic bonding among these bonds. Where the electronic charges are transferred from these cations to their neighboring anions due to the large electronegativity difference between $Ce^{3+}$ and $O^{2-}$.

Similarly, the contour lines of electronic charge density for the linear bonds ($O^{2-}$–$Tm^{3+}$–$O^{2-}$) distribute also in spherical shape to form the octahedra [$Tm^{3+}O^{2-}_6$], this indicates the presence of purely ionic bonds and electronic charges are transferred from $Tm^{3+}$ to $O^{2-}$. But the outer shape of some contour lines for ($O^{2-}$–$Tm^{3+}$–$O^{2-}$) changes from nearly spherical to oval shape, which indicates the formation of week ionic bonds. Here, the radius of outer spherical contour lines for anions $O^{2-}$ are larger than that for cations $Tm^{3+}$ in three compounds. In all crystal networks, since there is no overlapping between the contour lines around the ions, the bonding pattern among the cations $Ce^{3+}$ within octahedra [$Tm^{3+}O^{2-}_6$] is ionic. Thus, the nature of chemical bonds in these perovskites $CeTmO_3$ [$Tm^{3+}$ = Sc, Ti, V] compounds is purely ionic within all LDA, PBE-GGA and WC-GGA functionals. Fig. 12 (a-c) illustrates the spin-up and spin-down 3D version along [100] plane of the total electronic charge density for $CeTmO_3$ [$Tm^{3+}$ = Sc, Ti, V]. From these plots, some features such as ionic positions, spatial localization, spin-densities, orbitals symmetry, negative spin-density in interstitial sites, can be obviously recognized via their charges peaks. Also, the transfer charges can be evaluated by the difference between peaks around $Ce^{3+}$, $Tm^{3+}$ and $O^{2-}$ sites in spin-up, 45 – 70, and spin-down, 12 – 14, 3D electronic charge density.

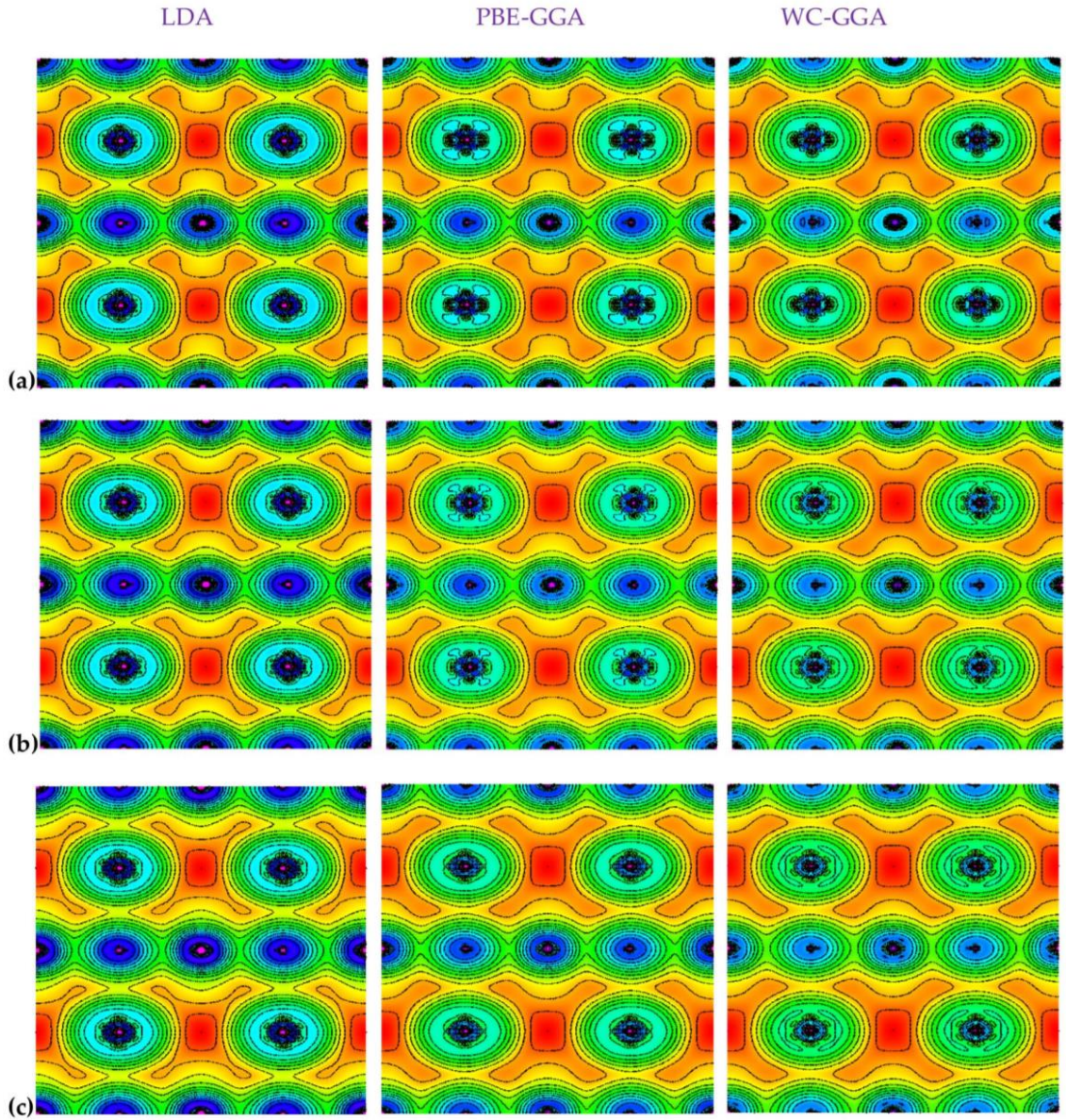

**Fig. 11.** Spin-up 2D total charge density along [100] plane of CeTmO$_3$ [Tm$^{3+}$ = Sc, Ti, V] obtained from LDA, PBE-GGA and WC-GGA functionals.

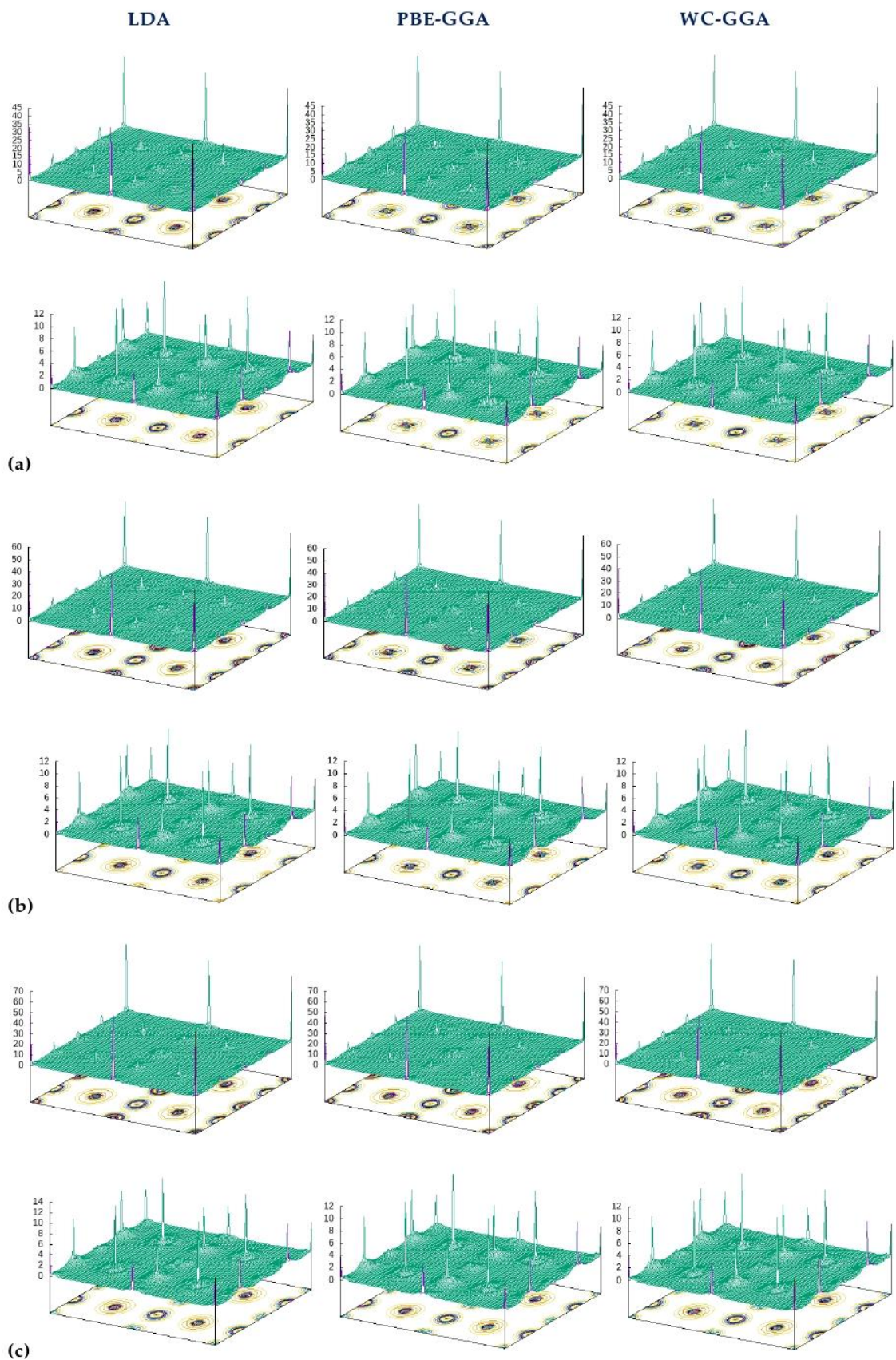

**Fig. 12.** Spin-up (up panels) and spin-dn (down panels) 3D total charge density along [100] plane of CeTmO$_3$ [Tm$^{3+}$ = (a) Sc, (b) Ti, (c) V] obtained from LDA, PBE-GGA and WC-GGA functionals.

## 4. Summary and Conclusions

By utilizing full-potential linear augmented plane-wave (FPLAPW) within the local density approximation (LDA) and generalized gradient approximation (GGA), under both Perdew-Burke-Ernzerh (PBE-GGA) and Wu-Cohen (WC-GGA) functionals, the main physical characteristics of $CeTmO_3$ compounds were systematically investigated. The detailed crystal, magnetic and electronic structures of cerium-based perovskites $CeTmO_3$ with [$Tm^{3+}$ = Sc, Ti, V] were computed by exploiting the Wien2k package based on the first-principles density functional theory (DFT). The optimized crystal structure parameters indicated that $CeTmO_3$ have cubic symmetry (space group Pm-3m, no. 221), with lattice constants agree with experimental values. All DFT results of spin magnetic moments, electronic band structures, density of states (DOS) and electronic charge density revealed that the three investigated $CeTmO_3$ compounds show ferromagnetic (FM) and metallic nature ($E_g$ = 0.0 eV), except for [$Tm^{3+}$ = Sc] compound, exhibits FM-half-metallic (HM) characteristics. In addition, the 2D and 3D plots for electronic charge density in [100] plane indicated the existence of ionic chemical bonding that dominant among all the energetic bonds ($O^{2-}$–$Ce^{3+}$–$O^{2-}$) and ($O^{2-}$–$Tm^{3+}$–$O^{2-}$) to construct the stable crystal structure in the perovskites $CeTmO_3$ compounds. Besides the effect of $Tm^{3+}$-site substitution [$Tm^{3+}$ = $Sc^{3+}$-$3d^0$, $Ti^{3+}$-$3d^1$, $V^{3+}$-$3d^2$] on the physical characteristics of $CeTmO_3$, utilizing LDA, PBE-GGA and WC-GGA functionals provided nearly similar results, where PBE-GGA is more realistic as compared to the other functionals.

## Acknowledgements

The author is would like to thank the Deanship of Scientific Research, Qassim University for financial support and funding publication of this project.